\documentstyle[twoside,fleqn,npb,epsfig]{article}
\newcommand{\insertfig}[2]{\mbox{\epsfxsize=#1cm \epsfbox{#2.eps}}}

\newcommand{\cA}{{\cal A}}

\newcommand{\cD}{{\cal D}}
\newcommand{\cK}{{\cal K}}

\newcommand{\cO}{{\cal O}}

\newcommand{\cP}{{\cal P}}

\newcommand{\cE}{{\cal E}}

\font\cmss=cmss12 
\def\1{\hbox{{1}\kern-.25em\hbox{l}}}
\def\bfZ{\relax{\hbox{\cmss Z\kern-.4em Z}}}

\begin{document}

\begin{titlepage}

\centerline{\large \bf Scaling violations and off-forward parton
                       distributions:}
\centerline{\large \bf leading order and beyond.}

\vspace{10mm}

\centerline{\bf A.V. Belitsky\footnote{Alexander von Humboldt Fellow.},
                D. M\"uller}

\vspace{10mm}

\centerline{\it Institut f\"ur Theoretische Physik, Universit\"at
                Regensburg}
\centerline{\it D-93040 Regensburg, Germany}

\vspace{30mm}

\centerline{\bf Abstract}

\vspace{0.8cm}

We give an outline of a formalism for the solution of the evolution
equations for off-forward parton distributions in leading and
next-to-leading orders based on partial conformal wave expansion and
orthogonal polynomials reconstruction.

\vspace{30mm}

\centerline{\it Talk given at the}
\centerline{\it 7th International Workshop on Deep Inelastic
                Scattring and QCD}
\centerline{\it DESY-Zeuthen, April 19-23, 1999}

\end{titlepage}

\title{Scaling violations and off-forward parton distributions:
       leading order and beyond}

\author{A.V. Belitsky\thanks{Alexander von Humboldt Fellow.}
	and D. M\"uller\address{Institut f\"ur Theoretische Physik,
                              Universit\"at Regensburg, \\
                              D-93040 Regensburg, Germany}}

\begin{abstract}

We give an outline of a formalism for the solution of the evolution
equations for off-forward parton distributions in leading and
next-to-leading orders based on partial conformal wave expansion and
orthogonal polynomials reconstruction.

\end{abstract}

\maketitle

The off-forward parton distributions (OFPD) are new non-perturbative
inputs used in exclusive electroproduction processes, like the hard
diffractive production of mesons and the deeply virtual Compton
scattering (DVCS), to parametrize hadronic substructure
\cite{DVCS,GuiVan98}. Their characteristic feature is a non-zero
momentum transfer in the $t$-channel which results into different
momentum fractions of constituents inside hadron.

The leading order amplitude, e.g.\ for the unpolarized DVCS, in
Leipzig-Ji's conventions used throughout \cite{DVCS}, looks like
\begin{equation}
\label{DVCSamplitude}
\cA \propto\!
\int_{- 1}^{1}\!\!\! dx \left\{ \frac{1}{x - \eta + i 0}
+ \frac{1}{x + \eta - i0} \right\} {\cal O} (x, \eta) ,
\end{equation}
with the quark OFPD ${\cal O} (x, \eta)$ given as the Fourier transform
of the light-cone string operator (in the light-cone gauge $B_+ = 0$)
\begin{equation}
\label{OFPD}
{\cal O} (x, \eta) = 2\! \int\frac{d \lambda}{2 \pi} e^{i \lambda x}
\langle p_2 | \bar\psi( - \lambda n ) \gamma_+ \psi( \lambda n )
| p_1 \rangle ,
\end{equation}
with the skewedness $\eta \equiv (p_1 - p_2)_+$ and the constraints
$x \in [-1, 1]$ coming form the support properties of the matrix
element (\ref{OFPD}). Several peculiar properties can be learned from
these: i) Translating the perturbative arguments used in the proof of
factorization formula (\ref{DVCSamplitude}) to non-perturbative domain
we immediately see that the $\cA$ exists provided the ${\cal O}
(\pm\eta, \eta)$ is continuous. ii) In different kinematical
regions of the phase space OFPD's share common properties with the
usual forward parton densities ($|x| > \eta$) and exclusive
distribution amplitudes ($|x| < \eta$), and thus hybrids in this sense.
iii) The amplitude (\ref{DVCSamplitude}) manifests Bjorken scaling.

The last property is violated once the QCD radiative corrections are
taken into account. The $Q$-dependence of the amplitude appears via
the scale dependence of the OFPD's which obey the generalized
evolution equation \cite{DVCS,BluGeyRob99,FraStr97,Beletal98}
\begin{equation}
\label{EvEq}
\frac{d}{d\ln Q^2}
{\mbox{\boldmath$\cal O$}} ( x, \eta )
= \int_{- 1}^{1} \! d x' \,
{\mbox{\boldmath$K$}} ( x, x', \eta )
{\mbox{\boldmath$\cal O$}} ( x', \eta ),
\end{equation}
with the kernel given by the series in the coupling constant $\alpha_s$.
The diagonalization of leading order kernel can be achieved exploiting
the consequences of conformal invariance of QCD at classical level.
The eigenstates of one-loop non-forward evolution equation are given
by Gegenbauer polynomials, $C_j^\nu$, --- with the numerical value of
the index $\nu$ depending on the parton species, --- which form an
infinite dimensional irreducible representation of the conformal group
in the space of bi-linear composite operators. Starting from two loop
order the conformal operators start to mix and the simple pattern of
one-loop evolution is broken so that the eigenfunctions generalize to
non-polynomial functions. In the basis of leading order conformal waves
\begin{equation}
\int_{- 1}^{1}\!\!\!\! d x \,
C_j^\nu\!\! \left( \frac{x}{\eta} \right)\!
{\mbox{\boldmath$K$}} ( x, x', \eta )
= - \frac{1}{2} \sum_{k = 0}^{j} \!\!
\mbox{\boldmath$\gamma$}_{jk}
C_k^\nu\!\! \left( \frac{x'}{\eta} \right)\!\!\!
\end{equation}
the anomalous dimension matrix is not diagonal and possesses
non-diagonal entries
\begin{equation}
\mbox{\boldmath$\gamma$}_{jk}
=
\mbox{\boldmath$\gamma$}^{\rm D}_{j} \delta_{jk}
+
\mbox{\boldmath$\gamma$}^{\rm ND}_{jk} ,
\quad\mbox{with}\quad
\mbox{\boldmath$\gamma$}^{\rm ND}_{jk} \propto \cO (\alpha_s^2).
\end{equation}

The disadvantage of the standard approach to the study of scaling
violation beyond leading order is the proliferation of Feynman graphs
required for calculation of $\mbox{\boldmath$\gamma$}^{(1)}_{jk}$. Our
approach which allows for an extremely concise analytical solution of
the problem \cite{BelMul98} is mainly based on four major observations:
i) The triangularity of the anomalous dimension matrix
$\mbox{\boldmath$\gamma$}_{jk}$ implies that its eigenvalues are
given by the diagonal elements and coincide with the well-known forward
anomalous dimensions. ii) Tree-level special conformal invariance implies
diagonal leading order matrix. One-loop violation of the symmetry induces
non-diagonal elements, thus, one-loop special conformal anomaly will
generate two-loop anomalous dimensions. iii) Scale Ward identity for
the Green function with conformal operator insertion coincides with
the Callan-Symanzik equation for the latter and thus the dilatation
anomaly is the anomalous dimension of the composite operator
$[\mbox{\boldmath$\cal O$}_j] [\int\! dx\, {\mit\Theta}_{\mu\mu}]
\propto \frac{1}{\epsilon} \sum_{k = 0}^{j} \mbox{\boldmath$\gamma$}_{jk}
[\mbox{\boldmath$\cal O$}_k]$. iv) The four-dimensional conformal algebra
provides a relation between the anomalies of dilatation and special
conformal transformations via the commutator $[\cD, \cK_-] = i \cK_-$.
Using these ideas we have deduced the form of the two-loop non-diagonal
elements to be \cite{BelMul98}
\begin{equation}
\mbox{\boldmath$\gamma$}^{{\rm ND}(1)} \!
=
[ \mbox{\boldmath$\gamma$}^{{\rm D}(0)}, \mbox{\boldmath$d$}\,
( \beta_0 - \mbox{\boldmath$\gamma$}^{{\rm D}(0)} )
+ \mbox{\boldmath$g$} ]_- ,
\end{equation}
where $\mbox{\boldmath$d$}$ is a simple matrix,
$\mbox{\boldmath$\gamma$}_{j}^{{\rm D}(0)}$ are the LO anomalous
dimensions of the conformal operators and $\beta_0$ is the one-loop
QCD Gell-Mann-Low function responsible for the violation of scale
invariance. The most nontrivial information about the special conformal
symmetry breaking is concentrated in the $\mbox{\boldmath$g$}$-matrices
which are residues of the special conformal symmetry breaking
counterterms at one-loop order $[\mbox{\boldmath$\cal O$}_j]
[\int\! dx\, x_- {\mit\Theta}_{\mu\mu}] \propto \frac{\alpha_s}{\epsilon}
\sum_{k = 0}^{j} a(j,k) \left\{ \mbox{\boldmath$g$}_{jk} + \dots \right\}
[\mbox{\boldmath$\cal O$}_k]$.

Unfortunately the eigenfunctions of the evolution kernels cannot be used
as a basis for expansion of OFPD since they do not form a complete set
of functions outside the region $|x/\eta| > 1$ where, however, the OFPD's
are nonvanishing in general. Our approach\footnote{Several LO methods are
available on a market \cite{ManPilWei98,GolMar98,ManKiv99}.}
\cite{Beletal98} to the study of the scale dependence of the OFPD is
based on the expansion of the distribution in series w.r.t.\ the complete
set of orthogonal polynomials, $\cP_j (x)$, on the interval $- 1 \leq x
\leq 1$ to preserve the support properties of the functions in question
\begin{equation}
\label{PolySeries}
{\mbox{\boldmath$\cal O$}} ( x, \eta, Q )
= \sum_{j = 0}^{J_{\rm max}}
\widetilde{\mbox{\boldmath$\cal P$}}_j (x)
{\mbox{\boldmath$\cal M$}}_j ( \eta, Q ) ,
\end{equation}
where formally $J_{\rm max} = \infty$. The expansion coefficients
can be reexpressed in terms of eigenstates of the evolution equation
(\ref{EvEq}) according to
\begin{eqnarray}
\label{JacobiMom}
{\mbox{\boldmath$\cal M$}}_j ( \eta, Q )
\!\!\!&=&\!\!\! \sum_{k = 0}^{j}
{\mbox{\boldmath$E$}}_{jk} (\eta)
\sum_{l = 0}^{k} \eta^{k - l}
{\mbox{\boldmath$B$}}_{k l} (Q, Q_0) \nonumber\\
&\times&\!\!\!
\mbox{\boldmath$\cE$}_l (Q, Q_0)
\mbox{\boldmath$\cal O$}_l ( \eta, Q_0 ) ,
\end{eqnarray}
where ${\mbox{\boldmath$E$}}_{jk} (\eta)$ is the overlap integral
\cite{Beletal98} between the one-loop eigenfunctions, $C_j^\nu$,
and the polynomials $\mbox{\boldmath$\cal P$}_j$. The conformal
moments at a reference scale $Q_0$ are defined as
\begin{equation}
\mbox{\boldmath$\cal O$}_j ( \eta, Q_0 )
= \eta^j \int_{- 1}^{1} dx\, C_j^\nu \left( \frac{x}{\eta} \right)
\mbox{\boldmath$\cal O$} ( x, \eta, Q_0 ) .
\end{equation}
All scale dependence in Eq.\ (\ref{JacobiMom}) is extracted to the
usual evolution operator which obeys the standard first order
differential equation
\begin{equation}
\frac{d}{d\ln Q}
\mbox{\boldmath$\cE$}
+
\mbox{\boldmath$\gamma$}^{{\rm D}}
\mbox{\boldmath$\cal E$} = 0 .
\end{equation}
Besides there is an additional dependence on the hard momentum
transfer, $Q$, which appears due to the mixing of the conformal
operators between themselves in two-loop approximation. This dependence
is governed by a new evolution equation of the form \cite{BelMul98}
\begin{equation}
\frac{d}{d\ln Q}
\mbox{\boldmath$B$}
+
\left[
\mbox{\boldmath$\gamma$}^{{\rm D}} ,
\mbox{\boldmath$B$}
\right]_-
+
{\mbox{\boldmath$\gamma$}}^{{\rm ND}}
\mbox{\boldmath$B$} = 0 .
\end{equation}

\begin{figure}[htb]
\unitlength1mm
\vspace{0.1cm}
\begin{center}
\begin{picture}(70,60)(0,0)
\put(0,0){\insertfig{6.4}{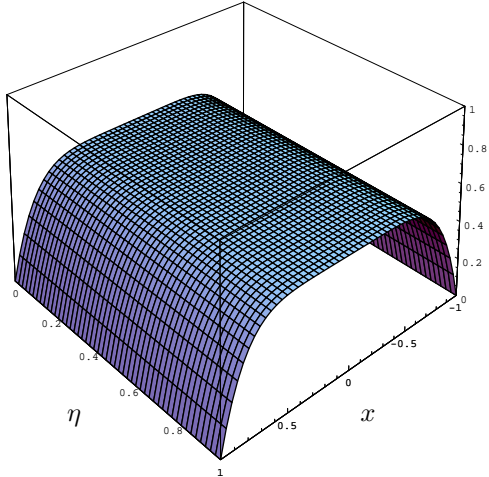}}
\put(8,8){$\eta$}
\put(47,8){$x$}
\end{picture}
\end{center}
\vspace{-1.1cm}
\caption{Sample $\eta$-independent valence-like input distribution
$\cO^{\rm val} (x, \eta) = \frac{7}{12} (1 - x^6)$ at low scale
$Q_0^2 = 0.2\, {\rm GeV}^2$.}
\label{Input}
\vspace{-0.8cm}
\end{figure}
\begin{figure}[htb]
\unitlength1mm
\vspace{0.1cm}
\begin{center}
\begin{picture}(70,60)(0,0)
\put(0,0){\insertfig{6.4}{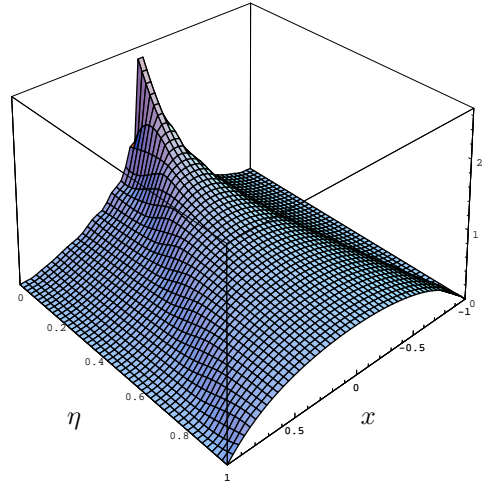}}
\put(8,8){$\eta$}
\put(47,8){$x$}
\end{picture}
\end{center}
\vspace{-1.1cm}
\caption{Distribution from Fig.\ \ref{Input} evolved with LO
formulae up to $Q^2 = 100\, {\rm GeV}^2$.}
\label{Q100}
\vspace{-0.8cm}
\end{figure}
\begin{figure}[htb]
\unitlength1mm
\vspace{-0.9cm}
\begin{center}
\begin{picture}(70,60)(0,0)
\put(8,0){\insertfig{5}{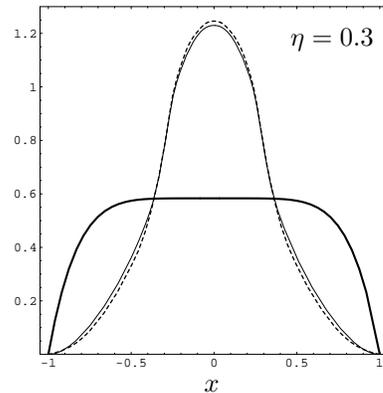}}
\put(33.5,-2){$x$}
\put(45,44){$\eta = 0.3$}
\end{picture}
\end{center}
\vspace{-1.2cm}
\caption{The input distribution $\cO^{\rm val}$ (thick curve) evolved at
LO (thin curve) and NLO (dashed curve) for $\eta = 0.3$ up to $Q^2 = 100\,
{\rm GeV}^2$.}
\label{Cut}
\vspace{-0.7cm}
\end{figure}

Making use of these results we are in a position to study the evolution
of OFPD explicitly. In order to save place let us address the non-singlet
sector only. The rough features of the shape of OFPD can be gained from
the perturbation theory itself. Assuming skewednessless input shown
in Fig.\ \ref{Input} at a very low normalization point typical for
phenomenological models of confinement we evolve it (with $J_{\rm max}
= 80$) upwards with momentum scale $Q^2 = 100\, {\rm GeV}^2$ (Fig.\
\ref{Q100}). The relative size of next-to-leading order effects is shown
for given $\eta$ in Fig.\ \ref{Cut}. Clearly, NLO effects do not exceed
the level of few percent in the non-singlet sector.

The specific feature of the evolution of the OFPD is that the partons
with momentum fractions $\eta < |x| < 1$ tend to penetrate into the
ER-BL-type region and once they do it they never return back from the
domain $x \in [-\eta, \eta]$.

{\bf Acknowledgements.}  A.B. was supported by the Alexander von
Humboldt Foundation.

\end{document}